%% file: bernardin1.tex
\documentclass[graybox]{svmult}

\usepackage{mathptmx}       
\usepackage{helvet}         
\usepackage{courier}        
\usepackage{type1cm}        
\usepackage{graphicx}        
\usepackage{multicol}        
\usepackage[bottom]{footmisc}

\usepackage{amssymb, amsmath}
\newcommand\bx{{\mathbf x}}

\newcommand\bq{{\mathbf q}}
\newcommand\bp{{\mathbf p}}
\newcommand\be{{\mathbf e}}

\newcommand\R{{\mathbb R}}

\newcommand\Z{{\mathbb Z}}

\newcommand\cE{\mathcal E}

\newcommand\CE{{\mathbb E}_{\scriptscriptstyle{N,T}}}

\newcommand\To{\mathbb T}

\newcommand\mc \mathcal

\begin{document}

\title*{Thermal conductivity for a chain of anharmonic oscillators perturbed by a conservative noise }
\titlerunning{Thermal conductivity for a perturbed nonlinear chain} 
\author{C\'edric Bernardin}
\institute{C\'edric Bernardin \at
Universit\'e de Lyon, CNRS (UMPA)\newline
Ecole Normale Sup\'erieure de Lyon,\newline
46, all\'ee d'Italie,\newline
69364 Lyon Cedex 07 - France.\newline \email{cbernard@umpa.ens-lyon.fr}}
%
%

\maketitle

\abstract*{We consider $d$-dimensional chains of (an)harmonic oscillators we perturb by a noise conserving energy or energy and momentum. We review the thermal conduction properties we obtained for these systems (\cite{BO}, \cite{BBO1}, \cite{B2}, \cite{BBO2}) and conclude by several open questions.  }

\abstract{We consider $d$-dimensional chains of (an)harmonic oscillators we perturb by a noise conserving energy or energy and momentum. We review the thermal conduction properties we obtained for these systems (\cite{BO}, \cite{BBO1}, \cite{B2}, \cite{BBO2}) and conclude by several open questions.  }


\section{Introduction}

The derivation of phenomenological laws from a microscopic description of the matter is one of the goals of statistical mechanics. Among them Fourier's law is probably one of the simple: when a small gradient $\nabla T$ of temperature is applied to a material, in the steady state,  the energy flux $J$ is proportional to the gradient of temperature
\begin{equation*}
J=-\kappa \nabla T
\end{equation*} 
The proportionality coefficient $\kappa$ is called the thermal conductivity. Despite its simplicity and the interest it has in the physical and mathematical community the derivation of Fourier's law from a microscopic model remains one of the main open question of nonequilibrium statistical mechanics (\cite{LLP}, \cite{BLR}, \cite{D2}).

In insulating crystals heat is transported by lattice vibrations, and since the pioneering work of Debye, systems of coupled anharmonic oscillators have been used as microscopic models for heat
conduction. They are classical system of particles interacting through a nearest neighbour interaction potential $V$ and which are in an external potential $W$. The Hamiltonian ${\mc H}$ is given by
\begin{equation*}
{\mc H} = \sum_{x \in \Lambda} \left( \frac{|p_x|^2}{2 m_x} + W(q_x) +\sum_{|y-x|=1} V(q_x-q_y) \right), \quad \Lambda \subset \Z^d
\end{equation*} 
where $m_x, q_x,p_x$ denotes the masse, position and momentum of the particle with equilibrium position $x \in \Lambda$.

It is well known that harmonic chains, because of their infinitely many conserved quantities, have infinite conductivity and do not obey Fourier's law (\cite{LRR}). This because phonons can traverse ballistically along the chain. It is often expected that enough strong nonlinearity or disorder (like  the presence of random masses) causes scattering between phonons and should imply a sufficiently fast decay of correlations for heat current and hence a normal conductivity. A rigorous treatment of a nonlinear system, even the proof of the existence of the conductivity coefficient, seems to be out of reach of current mathematical techniques. In this context the understanding of the coupled effect of nonlinearity and disorder is a challenge. 

The situation is in fact more complex. In some low dimensional systems ($d \le 2$) anomalous thermal conductivity is observed numerically and experimentally in nanotubes technology. The anomalous conductivity in low dimension has attracted a lot of attention in the literature, and it has been suggested that conservation of momentum is an important ingredient. There is no agreement, theoretically and numerically, about the exact dependance of the conductivity with the size of the system (\cite{LLP}).

Hence it makes sense to look at simple models which incorporate the important features that one believes are necessary to see normal transport. The main difficulty in Hamiltonian dynamics with a large number of degrees of freedom is to show that they behave ergodically, e.g. that the only time invariant measures locally absolutely continuous w.r.t. Lebesgue measure are, for infinitely extended spatial uniform systems, of the Gibbs type. For some stochastic lattice gases it can be proven but it remains a challenging problem for Hamiltonian dynamics. Taking advantage of mathematical techniques developed in the hydrodynamic limits communauty we introduce hybrid models between purely Hamiltonian systems and purely stochastic models which remain mathematically tractable but are sufficiently close to realistic systems to reproduce at least qualitatively what is observed for these systems. We consider chains of oscillators perturbed by a stochastic noise conserving energy or energy and momentum. These stochastic perturbations are here to simulate (qualitatively) the effective (deterministic) ergodicity coming from the Hamiltonian dynamics. 

The paper is organized as follows. In section \ref{sec:models} we introduce a model of coupled oscillators perturbed by a noise: the first noise conserves only energy and the second one conserves energy and momentum. In section \ref{sec:lrt} we review linear response theory and Green-Kubo formula. Section \ref{sec:ec} is devoted to the study of a chain of oscillators with the noise conserving only energy. For the harmonic homogenous harmonic chain we show Fourier's law is valid and compute the conductivity. For anharmonic chains we provide lower and upper bounds. The effect of disorder is considered in subsection \ref{subsec:DHC}. Then we consider the energy-momentum conserving model in section \ref{sec:emc} and show that in the homogenous harmonic case a breakdown of Fourier's law holds for low dimensional momentum conserving systems. We provide also upper bounds for the conductivity in the anharmonic case. We conclude the paper by open question in section \ref{sec:open}.

\vspace{0,5cm}
\textbf{Notations} : The canonical basis of $\R^d$ is noted
$(e_1,e_2, \ldots,e_d)$ and the coordinates of a vector $u \in
\R^d$ are noted $(u^1,\ldots,u^d)$. Its Euclidian norm $|u|$  is
given by $|u|=\sqrt{(u^1)^2+ \ldots +(u^d)^2}$ and the scalar
product of $u$ and $v$ is $u \cdot v$.  

If $N$ is a positive integer, $\mathbb T_N^d$ denotes the $d$-dimensional 
discrete torus of length $N$ and we identify $x = x + kN e_j$
for any $j= 1,\dots,d$ and $k\in \mathbb Z$.  

If $F$ is a function from ${\mathbb Z}^d$ (or $\To_N^d$) into $\mathbb
R$ then the (discrete) gradient of $F$ in the direction $e_j$ is
defined by $(\nabla_{e_j} F)(x)= F(x +e_j) -F(x)$ and the
Laplacian of $F$ is given by $(\Delta F) (x)= \sum_{j=1}^d
\left\{F(x +e_j)+F(x-e_j)-2F(x)\right\}$.

\section{The models}
\label{sec:models}
In this section we introduce deterministic nonlinear chains on a multidimensional lattice perturbed by a stochastic noise. The stochastic perturbations are such that they exchange momentum between particles with a local random mechanism that conserves total energy or total energy and total momentum.  

\subsection{Closed system}
\label{subsec:cs}

We first consider the closed system with periodic boundary conditions. The atoms are labeled by $x \in \mathbb T_N^d$. Momentum of atom $x$ is $p_x \in {\R^d}$, its displacement from its equilibrium position is ${q_x} \in \R^d$ and its mass is $m_x >0$. The configuration space is given by $\Omega_N = \{ (q_x,p_x) \in {\mathbb R}^d \times {\mathbb R}^d; \; x \in {{\To}_N^d}\}$. 
The Hamiltonian is given by
\begin{equation}
  \label{eq:hamilt}
  \mathcal H_N =  \sum_{x \in \To_N^d}  \left[ \frac{|p_x|^2}{2 m_x} + W(q_x)
    + \cfrac{1}{2} \sum_{|y-x|=1} V(q_x - q_y) \right] . \nonumber 
\end{equation}

We assume  that $V$ and $W$ have the following form: 
$$ 
V(q_x - q_y) = \sum_{j=1}^d V_j( q_x^j - q_y^j), \qquad W(q_x) = \sum_{j=1}^d W_j(q_x^j).
$$
and that $V_j, W_j$ are smooth, non-negative and even. We call $V$ the interaction potential, and $W$ the pinning
potential. The case where $W=0$ will be called unpinned. 

In the sequel we will refer to the \textit{$(\alpha,\nu)$-harmonic case}:
\begin{equation}
\label{eq:harm}
V_j (r)=\alpha r^2, \quad W_j (q)=\nu q^2, \quad \alpha >0, \quad \nu \geq 0
\end{equation}
for which explicit and generic results can be obtained.

The generator ${\mathcal L}_N$ of the dynamics is defined by
\begin{equation*}
{\mathcal L}_N ={\mathcal A}_N +\gamma {\mathcal S}_N
\end{equation*}    
where ${\mathcal A}_N$ is the Liouville operator corresponding to the Hamiltonian ${\mathcal H}_N$ 
\begin{equation*}
\label{eq:gen1A}
    {\mathcal A}_N = \sum_x \left\{ \partial_{p_x} \mathcal H_N \cdot \partial_{q_x} -
     \partial_{q_x}\mathcal H_N \cdot  \partial_{p_x} \right\}
\end{equation*}
and ${\mathcal S}_N$ is the generator of the Markovian noise. The parameter $\gamma>0$ regulates the strength of the noise. It acts only on momenta and is local. It consists in an infinitesimal exchange of momenta preserving some conservation laws. The first conservation law we impose is the energy conservation. Since the noise acts only on momenta it is equivalent to require conservation of kinetic energy. This corresponds to the so-called \textit{ energy conserving noise}. If we require also the conservation of total momentum then we get a second noise we call the \textit{energy-momentum conserving noise}. Let us denote $\pi_x = p_x/{\sqrt{m_x}}$. In the homogeneous case $m_x=1$, $\pi_x =p_x$.

The two noises have the following form
\begin{equation*}
\label{eq:gen1S}
{\mathcal S}_N =\cfrac{1}{4}\sum_{i,j=1}^d \sum_{\substack{x,z \in \To_N^d,\\|x-z|=1}} (X^{i,j}_{x,z})^2
\end{equation*}
where $X_{x,z}^{i,j}$ is equal to
\begin{equation*}
X^{i,j}_{x,z}= \pi_{z}^j \partial_{\pi^i_{x}} - {\pi^i_{x}}\partial_{\pi^j_{z}}
\end{equation*}
for the energy conserving noise and equal to
\begin{equation*}
  \label{eq:Xfield}
   X^{i,j}_{x, z} = (\pi^j_z-\pi^j_x) (\partial_{\pi^i_z}
   - \partial_{\pi^i_x})  -(\pi^i_z-\pi^i_x) (\partial_{\pi^j_z}
   - \partial_{\pi^j_x}) .
\end{equation*}
for the energy-momentum noise if $d\ge 2$. If $d=1$ in order to conserve total momentum and total kinetic energy, we have to consider a random exchange of momentum between three consecutive atoms (\cite{BBO2}).

The interpretation of the vector fields is the following. To be specific we take $d=1$ and consider the energy conserving noise. In this case $X_x=X^{1,1}_{x,x+1} = (\pi_{x+1} \partial_{\pi_x} -\pi_x \partial_{\pi_{x+1}})$. Observe that $X_x$ is the vector field tangent to the circle $C_x=\{(\pi_x,\pi_{x+1}); \; \pi_x^2 +\pi_{x+1}^2 =1\}$ so that $X_{x}^2$ generates a diffusion on $C_x$. In fact it is nothing else than a standard Brownian motion on the circle. The generator ${\mc S}_N$ corresponds to a system of coupled Brownian diffusions preserving the kinetic energy $\sum_{x \in \To_N^d} |\pi_x|^2$. The energy-momentum conserving noise is defined by a similar procedure but the surface $C_x$  has to be replaced by the surface of constant kinetic energy and constant momentum. In dimension $1$, this surface is reduced to a point and it explains why we have to consider a three-body interaction.  

Because the noise conserves energy, a family of stationary translations invariant probability measures for ${\mc L}_N$ is given by the Gibbs measures. In the energy conserving case they are parametrized by inverse temperature $\beta=T^{-1}$ and in the energy-momentum conserving model by inverse temperature $\beta$ and mean momentum average $\bar p$. We denote the Gibbs measure with inverse temperature $\beta=T^{-1}$ and zero momentum average by $\mu_{\scriptscriptstyle{N,T}}$. It is given by
\begin{equation*}
\mu_{\scriptscriptstyle{N,T}}(d\bq \,d\bp)= Z_{\scriptscriptstyle{N,T}}^{-1} \exp(-\beta {\mc H}_N) d\bq \,d\bp
\end{equation*}    
where $Z_{\scriptscriptstyle{N,T}}$ is the partition function. Expectation with respect to $\mu_{\scriptscriptstyle{N,T}}$ is denoted by $\langle \cdot \rangle_{N,T}$. Remark that in ${\mathbb L}^2 (\mu_{\scriptscriptstyle{N,T}})$ the Hamiltonian vector field ${\mc A}_N$ is antisymmetric and the noise ${\mc S}_N$ is symmetric.

\subsection{Open system}
\label{subsec:os}

We now consider the case where the system is in contact with thermal baths at different temperatures $T_\ell$ and $T_r$.  Thermal baths are given by Ornstein-Uhlenbeck processes with the corresponding temperature. To simplify notations we take $d=1$. The configuration space is now given by $\chi_N= \{(p_x,q_x) \in {\mathbb R} \times {\mathbb R}; \; x =1, \ldots, N\}$. The generator of the evolution has the form
\begin{equation*}
  \begin{split}
    {\tilde {\mc L}}_N =& \sum_{x=1}^{N} \left\{ \partial_{p_x} {\tilde{\mathcal H}}_N \cdot \partial_{q_x} - \partial_{q_x}{\tilde{\mathcal H}}_N \cdot  \partial_{p_x} \right\} \\
& + \frac \gamma 2 \sum_{x=1}^{N-1} (X^{1,1}_{x, x +1})^2
+  \frac 12 \left(T_\ell \partial_{p_1}^2 - p_1 \partial_{p_1} \right)
+  \frac 12 \left(T_r \partial_{p_{N}}^2 - p_{N} \partial_{p_{N}} \right)
  \end{split}
\end{equation*}
We have to specify boundary conditions for ${\tilde{\mc H}}_N$. For example one can define
\begin{equation*}
{\tilde{\mc H}}_N =  \sum_{x=1}^{N-1}  \left[ \frac{|p_x|^2}{2 m_x} + W(q_x)
    + \cfrac{1}{2} \sum_{|y-x|=1} V(q_x - q_y) \right] . \nonumber 
\end{equation*}
with $q_0$ and $q_{N+1}$ fixed. 

Even if one can prove there exists a unique stationary probability measure $\langle \cdot \rangle_{\scriptscriptstyle{N,ss}}$ for the process, there is in general no formula to express it. The only case where one knows $\langle \cdot \rangle_{\scriptscriptstyle{N,ss}}$ is the equilibrium case $T_\ell=T_r=T$ where the stationary measure is given by the Gibbs measure at temperature $T$. Otherwise the probability measure  $\langle \cdot \rangle_{\scriptscriptstyle{N,ss}}$ is called a {\textit{nonequilibrium}} stationary state.

\section{Thermal conductivity and linear response theory}
\label{sec:lrt}
In this section we review briefly linear response theory and Green-Kubo formula for thermal conductivity. Derivation of the Green-Kubo formula is heuristic and even its (mathematical) existence is a challenging problem. Roughly speaking Green-Kubo formula is the space-time variance of the total current at equilibrium. Thermal conductivity is a transport coefficient defined by considering the system out of equilibrium. Linear response theory express the fact that if temperatures $T_\ell$ and $T_r$ are different but close then a linear approximation is valid and the thermal conductivity $\kappa (T)$ is equal to the Green-Kubo formula $\kappa^{GK} (T)$. Such a formula belongs to the family of \textit{fluctuation-dissipation} theorems since it relates dissipation (i.e. the thermal conductivity) to fluctuations (i.e. fluctuations of the total current).

Defining the energy of the atom $x$ as
\begin{equation*}
  \label{eq:energyx}
  {\cE}_{x} = \frac{1}{2m_x}  p_x^2 \, +\,  {W( q_x)} \,+\,
\cfrac{1}{2}\sum_{y: |y -x|=1}
V(q_{y} - q_x) 
\end{equation*}
the energy conservation law can be read locally as
\begin{equation*}
   {\cE}_{x}(t) -   {\cE}_{x}(0) =  \sum_{k=1}^d \left(\,
J_{{x-e_k, x}}([0,t]) - J_{{x,x +e_k}}([0,t])\,\right)
 \end{equation*}
where $J_{x,x +e_k}([0,t])$ is the total energy current
between $x$ and $x +e_k$ up to
time $t$. This can be written as
\begin{equation*}
  \label{eq:tc}
  J_{x, x +e_k}({[0,t]})=\int_0^t j_{x, x +e_k}(s) \; ds +
  M_{x, x +e_k}(t) 
\end{equation*}
In the above $M_{x, x +e_k}(t)$ are  martingales that can be
written explicitly as It\^o stochastic integrals.

The instantaneous energy
currents $j_{x,x +e_k}$ satisfy the equation 
\begin{equation*}
  {\mc L}_N {\cE}_{x} = \sum_{k=1}^d \left(
j_{x-e_k, x} - j_{x,x +e_k}\right)
\end{equation*}
and it can be written as
\begin{equation}
  \label{eq:1}
  j_{x, x +e_k} = j^{a}_{x, x + e_k} +\gamma j_{x, x +e_k}^s 
\end{equation}
The first term in (\ref{eq:1}) is the Hamiltonian contribution to the
energy current
\begin{equation*}
  \label{eq:21}
  \begin{split}
    j^a_{x,x +e_k} &= -\cfrac{1}{2} (\nabla V)(q_{x+e_k} -
    q_x)\cdot \left(\cfrac{\pi_{x+e_k}}{m^{1/2}_{x+e_k}} + \cfrac{\pi_x}{m^{1/2}_x}\right)\\
    &= -\frac 12 \sum_{j=1}^d V'_j(q^j_{x+e_k} -
    q^j_x) \left(\cfrac{\pi^j_{x+e_k}}{m^{1/2}_{x+e_k}} + \cfrac{\pi^j_x}{m^{1/2}_x}\right)
  \end{split}
\end{equation*}
while $j^s_{x,x+e_k}$ is the noise contribution. 

For the energy conserving noise we have
\begin{equation*}
j^s_{x,x+e_k} = -\cfrac{1}{d} \nabla_{e_k} |\pi_x|^2
\end{equation*}

In the energy-momentum conserving case, in $d\ge 2$, it is
\begin{equation*}
  \label{eq:31}
     j^s_{x,x +e_k} =- \nabla_{e_k} |\pi_x|^2
 \end{equation*}
and in $d=1$ is
\begin{equation*}
\begin{split}
   j^s_{x, x + 1} =& -\nabla \varphi (\pi_{x-1},\pi_x,\pi_{x+1})
  \\
 \varphi (\pi_{x-1},\pi_x,\pi_{x+1})&=
\frac 16 [\pi_{x+1}^2 + 4 \pi_{x}^2 +
    \pi_{x-1}^2 +  \pi_{x+1} \pi_{x-1} -2 \pi_{x+1} \pi_{x}-2 \pi_{x} \pi_{x-1}]
\end{split}
\end{equation*}

The particular form of $j^s$ is not very important. What is relevant is the fact that $j^s_{x,x+e_k}$ is a {\textit{discrete gradient}} $\nabla_{e_k}  \phi_x$ of a local function $\phi_x$. It means a weak form of Fourier's law is valid at the microscopic level as soon as we can relate $\phi_x$ to the (macroscopic) local temperature. Observe that it is not the case for the Hamiltonian part of the current and one of the main difficulties is to express $j^a$ as the sum of a discrete gradient and a small term (see (\ref{eq:fde})).

\subsection{Nonequilibrium setting}

We consider the system out of equilibrium (see subsection \ref{subsec:os}) in contact with two heat baths at different temperatures $T_\ell$ and $T_r$ in the first direction. The conductivity $\kappa (T)$ is defined by the thermodynamic limit
\begin{equation*}
\kappa (T) =\lim_{T_\ell, T_r \to T} \lim_{N \to \infty} \cfrac{\langle j_{0,e_1}\rangle_{\scriptscriptstyle{N,ss}}}{T_\ell-T_r}
\end{equation*}
We have seen in the introduction  one expects that for one and two dimensional systems conserving momentum such a limit is equal to infinity. In order to estimate this divergence one can study the finite size thermal conductivity
\begin{equation}
\label{eq:120}
\kappa_N (T) = \lim_{T_\ell, T_r \to T} \cfrac{\langle j_{0,e_1}\rangle_{\scriptscriptstyle{N,ss}}}{T_\ell -T_r}
\end{equation}
Under suitable conditions his quantity is well defined even for purely Hamiltonian chains but it is not straightforward. It is often expected that
$$\lim_{N \to \infty} \kappa_{N} (T) =\kappa (T)$$
but it is not obvious since there is an exchange of limits.

\subsection{Green-Kubo formula}
The difficulty arising in the study of $\kappa (T)$ is that we have no explicit representation of the nonequilibrium stationary state $\langle \cdot \rangle_{\scriptscriptstyle{N,ss}}$. Since $T_\ell$ and $T_r$ are close it is suggestive to use a perturbative approach to compute $\kappa (T)$. 

Performing a first order development in the stationary state $\langle \cdot \rangle_{\scriptscriptstyle{N,ss}}$ we get that
$$\kappa(T)=\kappa^{GK}(T)$$
where the Green-Kubo formula for the conductivity $\kappa^{GK} (T)$ is given by (\cite{Sp}, pp. 188--190) 
 \begin{equation}
  \label{eq:gcc}
 \kappa^{GK} (T) = \lim_{t\to \infty} \lim_{N\to \infty} 
   \frac 1{2 T^2 t} \sum_{x \in {\mathbb T}_N^d} \CE 
   \left[ \,J_{x,x+e_1}([0,t]) 
     J_{0,e_1}([0,t]) \, \right] 
\end{equation}
Here $\CE$ indicates the expectation with respect to the equilibrium dynamics starting with the Gibbs measure $<\cdot>_{\scriptscriptstyle{N,T}}$ at temperature $T$.  This definition itself is formal since we have to prove existence of the limits.
   
 By standard stochastic calculus and a time-reversal argument (\cite{BBO2}) one can establish the following equality
 \begin{equation}
 \label{eq:5c}
   \begin{split}
   & \frac 1{2 T^2 t} \sum_x \CE \left(
        J_{x,x+e_1}([0,t]) J_{0,e_1}([0,t]) \right)\\
   & = (2 T^2 N^d t)^{-1} \CE\left(\left[\sum_x \int_0^t
        j^a_{x,x+e_1}(s) ds\right]^2 \right) + \frac{\gamma}d
  \end{split}
\end{equation}

Here the term $\gamma/d$ is due to the presence of the noise. This is the first term which is of interest. In view of the Green-Kubo formula the anomalous behavior of the conductivity should appear in a slow time-decay of the time correlation of the (Hamiltonian part) of the current. 
   
Let us denote
\begin{equation*}
{\frak J}_{e_1} = \sum_{x \in {\To}_N^d} j^{a}_{x,x+e_1}
\end{equation*}
In order to study the large time behavior of 
\begin{equation*}
C (t)= \lim_{N \to \infty} C_N (t), \quad C_N (t)=(2 T^2 t N^d)^{-1}{\mathbb E}_{\scriptscriptstyle{N,T}} \left(\left[\int_0^t {\frak J}_{e_1} (s) ds\right]^2 \right)
\end{equation*}
we study the asymptotics as $N \to \infty$ and then $\lambda \to 0$ of the Laplace transform ${\frak L}_N (\lambda)$ of $ t C_N (t)$
\begin{equation*}
{\frak L}_N (\lambda)= \int_{0}^\infty e^{-\lambda t} t C_N (t) dt
\end{equation*}
By stationarity and integration by parts, we have
\begin{equation*}
{\frak L}_N(\lambda)=\cfrac{1}{\lambda^2 T^2} \int_0^\infty dt e^{-\lambda t} {\mathbb E}_{\scriptscriptstyle{N,T}} \left[{\frak J}_{e_1} (t) {\frak J}_{e_1} (0)\right]
\end{equation*} 
Denote by $e^{t {\mc L}_N}$ the semigroup generated by ${\mc L}_N$ and remark that ${\mathfrak J}_{e_1} (t) = e^{t {\mc L}_N} {\mathfrak J}_{e_1}$  then
\begin{equation}
\label{eq:16}
{\frak L}_N (\lambda)= \cfrac{1}{\lambda^2 T^2}  \langle {\mathfrak J}_{e_1}, (\lambda -{\mc L}_N)^{-1} {\mathfrak J}_{e_1} \rangle_{\scriptscriptstyle{N,T}}
\end{equation}

A normal finite conductivity corresponds (in a Tauberian sense) to a positive finite limit of $\lambda^{2} {\frak L}_N (\lambda)$ as $N \to \infty$ and then $\lambda \to 0$. In this case, the conductivity $\kappa (T)$ should be equal to the following form of the Green-Kubo formula 
\begin{equation}
\label{eq:18}
\kappa^{GK} (T)  =\gamma/d + \lim_{\lambda \to 0} \lim_{N\to \infty} N^{-d} \langle {\mathfrak J}_{e_1}, (\lambda -{\mc L}_N)^{-1} {\mathfrak J}_{e_1} \rangle_{\scriptscriptstyle{N,T}}
\end{equation}

No general argument gives the existence of the limits in (\ref{eq:gcc}) and in (\ref{eq:18}) nor that if they exist they are equal. 

Observe that Green-Kubo formula (\ref{eq:gcc}) predicts only the value of the thermal conductivity $\kappa (T)$ (defined in the nonequilibrium setting). If the thermal conductivity is infinite it says a priori nothing about the behavior of the finite size thermal conductivity $\kappa_N (T)$ defined by (\ref{eq:120}). To overcome this problem we define the \textit{truncated Green Kubo formula} by
\begin{equation*}
\kappa_N^{GK} (T) =  (2 T^2 N^d t_N)^{-1} \CE\left(\left[\sum_\bx \int_0^{t_N}
        j^a_{\bx,\bx+\be_1}(s) ds\right]^2 \right) + \frac{\gamma}d
\end{equation*}
where $t_N=N/v_s$ with $v_s$ the sound velocity defined by
\begin{equation*}
 v_s=\lim_{k\to 0} |\partial_{k^1} \omega(k)|
\end{equation*}
and where 
$$\omega(k)= \left( \sum_{j=1}^d {W_{j}''} (0) +4{V_{j}}'' (0) \sin^2 (\pi k^j)\right)^{1/2}$$ 
is the dispersion relation of the approximated linear system. This definition of the conductivity of the finite system is motivated by the following consideration: in the harmonic case the finite size thermal conductivity $\kappa_N (T)$ can be obtained by this truncation technique (in a rigorous way) and we expect this is still valid for the anharmonic chain. In the linear interactions approximation $\nabla_{k}\omega(k)$ is the group velocity of the $k$-mode waves, which are the heat carriers, and typically $v_s$ is an upper bound for these velocities. Consequently $t_N$ is the typical time a low $k$ (acoustic) mode takes to cross around the system once (see\cite{LLP}). Typically $v_s$ is of order one and we will take $v_s=1$ in the sequel.

\section{Energy conserving model}
\label{sec:ec}
In this section we state the results obtained for the energy conserving noise. As expected one can not obtain infinite conductivity for this model since momentum is not conserved. In the homogenous harmonic case one computes explicitly the Green-Kubo formula $\kappa^{GK} (T)$ and also the conductivity $\kappa(T)$ in the one-dimensional unpinned system. It turns out that $\kappa(T)=\kappa^{GK} (T)$ so that predictions of linear response theory is valid although the proof does non use perturbative arguments. In the anharmonic homogenous case one can establish lower and upper bounds for $\kappa^{GK} (T)$ indicating a positive finite conductivity. The proof of the convergence of the Green-Kubo formula in the anharmonic case remains open.   
We are also interested in the effect of disorder (random masses) in the energy conserving model. This is the contain of subsection \ref{subsec:DHC}.

\subsection{Homogenous harmonic chain}
Here we consider the homogenous  $(\alpha,\nu)$-harmonic case (\ref{eq:harm}) with all masses equal to $1$.

Let us define 
\begin{equation*}
\label{eq:D}
D = \int_{\xi \in [0,1]^d} \left( \cfrac{4\alpha^2 \sum_{j=1}^d \sin^2 (\pi \xi^j)}{\nu +4\alpha \sum_{j=1}^d \sin^2 (\pi \xi^j)}\right)d\xi^1 \ldots d\xi^d
\end{equation*}
We have the following theorem

\begin{theorem}
$\kappa^{GK}(T)$ defined by (\ref{eq:gcc}) is finite (pinned or unpinned) in any dimension and given by
\begin{equation*}
\kappa^{GK}(T) = \cfrac{D}{\gamma} +\cfrac{\gamma}{d}
\end{equation*}
\end{theorem}

\begin{proof}
Recall (\ref{eq:16}). A simple but crucial computation shows that
\begin{equation*}
(\lambda -{\mc L}_N)^{-1} {\frak J}_{e_1}= \cfrac{{\frak J}_{e_1}}{\lambda +\gamma}
\end{equation*} 
Let $D_N=D_N (\alpha,\nu)$ be the constant
\begin{equation*}
\label{eq:D_N}
D_N= T^{-1} \alpha^2 \sum_{k=1}^d \langle (q_{e_1}^k -q_{-e_1}^k)^2\rangle_{\scriptscriptstyle{N,T}}=\cfrac{1}{N^d} \sum_{\xi \in \To_N^d} \left( \cfrac{4\alpha^2 \sum_{j=1}^d \sin^2 (\pi \xi^j /N)}{\nu +4\alpha \sum_{j=1}^d \sin^2 (\pi \xi^j /N)}\right)
\end{equation*}
One computes easily $\lim_{N \to \infty}  \langle {\frak J}_{e_1}, {\frak J}_{e_1} \rangle_{N,T}$ and after inversion of the Laplace transform, one gets
\begin{equation*}
C_N (t)=\cfrac{D_N}{\gamma} \left( 1+ \cfrac{1}{\gamma t} (1-e^{-\gamma t}) \right) 
\end{equation*}
Then theorem follows.
\end{proof}

Observe that if the noise becomes weaker (i.e. $\gamma \to 0$), we obtain a purely homogenous harmonic chain and the thermal conductivity is infinite. 

In the following theorem we study the one-dimensional unpinned system in contact with two heat baths at temperature $T_\ell$ and $T_r$. We show that conductivity is finite and coincides with the Green-Kubo formula. The proof remains valid in any dimension and with or without pinning as soon as we are able to prove the following bound
\begin{equation}
\label{eq:22}
\forall x \in \{1,\ldots,N\}, \quad \langle e_x \rangle_{\scriptscriptstyle{N,ss}} \le C
\end{equation}
where $C>0$ is independent of $N$. Unfortunately it has been proved only in the one dimensional unpinned case.
 
\begin{theorem}{\cite{BO}}
\label{th:FL}
Consider the one dimensional $(\alpha,\nu)$-harmonic case (\ref{eq:harm}) with $\nu=0$. For any $\gamma > 0$
  \begin{equation}
    \label{eq:fourier}
    \lim_{N\to\infty} N  <j_{x,x+1}> = \alpha \left(\gamma +\gamma^{-1}\right)\left(T_\ell - T_r\right)
  \end{equation}
 Hence we have $\kappa (T)=\kappa^{GK} (T)= \alpha (\gamma +\gamma^{-1})$
\end{theorem}

The proof of this theorem can be found in \cite{BO}. It is based on entropy production bounds and use an explicit decomposition of the current $j_{x,x+1}$ as the sum 
\begin{equation}
\label{eq:fde}
j_{x,x+1}=\nabla\phi_x +{\mathcal L}_N h_x
\end{equation}
where $h_x, \phi_x$ are two explicit local functions and $\nabla$ is the discrete gradient. Hence the current is the sum of a dissipative part (a spatial gradient) and a fluctuating part (a time derivative).  For this reason we call this equation a \textit{microscopic fluctuation-dissipation relation} (\cite{B1},\cite{B3},\cite{BO}).

\subsection{Homogenous anharmonic chain}
\label{subsec:ac}
The introduction of nonlinearity in the Hamiltonian dynamics complicates considerably the problem. We do not have any proof of the existence of $\kappa^{GK}(T)$ nor $\kappa (T)$. At least we have estimates which indicate that a finite strictly positive conductivity is expected. By strictly positive we mean that the Hamiltonian contribution to the Green Kubo formula (the second term in (\ref{eq:18})) is strictly positive, the term $\gamma/d$ being only due to the noise and of no interest for the study of conduction properties of the underlying deterministic dynamics. We consider the (formal) Green-Kubo formula (\ref{eq:18}).

\begin{proposition}
 Assume that
\begin{equation*}
N^{-d} \langle {\mathfrak J}_{e_1}, {\mathfrak J}_{e_1} \rangle_{\scriptscriptstyle{N,T}} \le C
\end{equation*}
with a constant $C$ independent of $N$.  Then there exists a positive constant $C'$ independent of $N$ such that for any $\lambda >0$ and $N$,
\begin{equation}
N^{-d} \langle {\mathfrak J}_{e_1}, (\lambda -{\mc L}_N)^{-1} {\mathfrak J}_{e_1} \rangle_{\scriptscriptstyle{N,T}} \le C'
\end{equation}
\end{proposition}

Observe that the assumption done in this proposition is natural and is satisfied for reasonable potentials $V,W$ (see (\ref{eq:JJ})). This shows that if Green-Kubo formula (\ref{eq:18}) converges then $\kappa^{GK}(T)$ is finite.

\begin{proof}
Introduce the following so-called $H_{1,\lambda}$ and $H_{-1,\lambda}$ norms defined by
\begin{equation*}
\| f\|^2_{\pm 1,\lambda} = \langle f, (\lambda -\gamma {\mathcal S}_N)^{\pm} f \rangle_{\scriptscriptstyle{N,T}}
\end{equation*}

We have the following variational formula (\cite{B2})
\begin{equation*}
 \langle {\mathfrak J}_{e_1}, (\lambda -{\mc L}_N)^{-1} {\mathfrak J}_{e_1} \rangle_{\scriptscriptstyle{N,T}}
 =\sup_{u} \left\{ 2\langle {\mathfrak J}_{e_1}, u\rangle_{\scriptscriptstyle{N,T}} - \|u \|^2_{1,\lambda} -\| {\mc A}_N u \|^2_{-1,\lambda} \right\}
\end{equation*}
To obtain the upper bound we forget the term $\| {\mc A}_N u\|_{-1,\lambda}^2$ and we get
\begin{equation*}
 \langle {\mathfrak J}_{e_1}, (\lambda -{\mc L}_N)^{-1} {\mathfrak J}_{e_1} \rangle_{\scriptscriptstyle{N,T}} \le \sup_{u} \left\{ 2\langle {\mathfrak J}_{e_1}, u\rangle_{\scriptscriptstyle{N,T}} - \|u \|^2_{1,\lambda} \right\}
\end{equation*}
It is well known that
\begin{equation*}
 \sup_{u} \left\{ 2\langle {\mathfrak J}_{e_1}, u\rangle_{\scriptscriptstyle{N,T}} - \|u \|^2_{1,\lambda} \right\}=\langle {\mathfrak J}_{e_1}, (\lambda -\gamma{\mc S}_N)^{-1} {\mathfrak J}_{e_1}  \rangle_{\scriptscriptstyle{N,T}}
\end{equation*}
and a simple computation shows that
\begin{equation*}
(\lambda -\gamma {\mc S}_N)^{-1} {\frak J}_{e_1}= \cfrac{{\frak J}_{e_1}}{\lambda +\gamma}
\end{equation*}
We are left to prove that 
\begin{equation*}
 N^{-d} \langle {\mathfrak J}_{e_1}, {\mathfrak J}_{e_1} \rangle_{\scriptscriptstyle{N,T}} \le C
\end{equation*}
with a constant $C$ independent of $N$. It is exactly our assumption.
\end{proof}

We expect also a lower bound of the form 
\begin{equation*}
C^{-1} \le N^{-d} \langle {\mathfrak J}_{e_1}, (\lambda -{\mc L}_N)^{-1} {\mathfrak J}_{e_1} \rangle_{\scriptscriptstyle{N,T}} 
\end{equation*}
with $C>0$ independent of $\lambda$ and $N$.

 The strategy to prove such a lower bound is straightforward: we have to find a good test function $v_{\scriptscriptstyle{N,\lambda}}$ and show that
\begin{equation*}
N^{-1} \left\{ 2\langle {\mathfrak J}_{e_1}, v_{\scriptscriptstyle{N,\lambda}}\rangle_{\scriptscriptstyle{N,T}} - \|v_{\scriptscriptstyle{N,\lambda}} \|^2_{1,\lambda} -\| {\mc A}_{\scriptscriptstyle{N}} v_{\scriptscriptstyle{N,\lambda}} \|^2_{-1,\lambda} \right\}  \ge C^{-1}
\end{equation*}

Unfortunately we are not able to prove this lower bound for general $V$ and $W$ but only for unpinned systems ($W=0$) in $d=1$. We note $r_x = q_{x+1} -q_x$. Then the $p$'s and the $r$'s are independent variables under the Gibbs measure $\langle \cdot \rangle_{\scriptscriptstyle{N,T}}$. Some conditions on $V$ are imposed
\begin{equation*}
{\text{Var}}(V'' (r_0)) \le C, \quad {\text{Var}} (V' (r_0)) \le C
\end{equation*}
with $C>0$ and ${\text {Var}} (F(r_0))$ is the variance (independent of $N$) of $F(r_0)$ under the Gibbs measure $\langle \cdot \rangle_{\scriptscriptstyle{N,T}}$.

\begin{proposition}
Under the conditions above there exists a positive constant $C'$ independent of $N$ and $\lambda$ such that
 \begin{equation*}
C' \le N^{-1} \langle {\mathfrak J}_{e_1}, (\lambda -{\mc L}_N)^{-1} {\mathfrak J}_{e_1} \rangle_{\scriptscriptstyle{N,T}} 
\end{equation*}
\end{proposition}

\begin{proof}

We choose 
$$v_{\scriptscriptstyle{N,\lambda}} = -a\sum_{x} p_x (V' (r_x) + V' (r_{x-1}))=2a {\mathfrak J}_{e_1}$$
with $a>0$ we will precize later. We have
\begin{equation*}
{\mc A}_N v_{\scriptscriptstyle{N,\lambda}}=  -a\sum_x p_x^2 \left( V''(r_{x-1})-V'' (r_x) \right)
\end{equation*}  

Let $G_{\lambda, N} (z)$ the solution on $\To_N$ of 
$$ (\lambda -\gamma \Delta) G_{\scriptscriptstyle{N,\lambda}} =\delta_0 (\cdot) $$
Observe that ${\mc S}_N (p_x^2)= \Delta (p_x^2)$ then we get
\begin{equation*}
(\lambda -\gamma {\mc S}_N )^{-1} ({\mc A}_n v_{\scriptscriptstyle{N,\lambda}}) = -a \sum_{x,z} G_{\scriptscriptstyle{N,\lambda}} (x-z) p_z^2 \left( V'' (r_{x-1})- V'' (r_x)\right) 
\end{equation*}
It follows that
\begin{equation*}
\| {\mc A}_N v_{\scriptscriptstyle{N,\lambda}} \|_{-1,\lambda}^2 = a^{2} \sum_{x,y,z} G_{\lambda, N} (x-z) \langle p_y^2 p_z^2 \left( V'' (r_{y-1}) -V'' (r_y)\right) \left( V'' (r_{x-1}) -V'' (r_x)\right)\rangle_{\scriptscriptstyle{N,T}} 
\end{equation*}
Observe now the Gibbs measure $\langle \cdot \rangle_{\scriptscriptstyle{N,T}}$ is product. Then an easy computation shows
\begin{equation*}
\| {\mc A}_N v_{\scriptscriptstyle{N,\lambda}} \|_{-1,\lambda}^2 = N a^2 T {\text {Var}} (V''(r_0)) (\Delta G_{\scriptscriptstyle{N,\lambda}})(0)
\end{equation*}
By discrete Fourier transform we have
\begin{equation*}
(\Delta G_{\scriptscriptstyle{N,\lambda}})(0) =\cfrac{1}{N} \sum_{k \in \To} \cfrac{4 \sin^{2}(\pi k/N)}{\lambda +4\gamma \sin^2 (k \pi /N)} \le \gamma^{-1}
\end{equation*}
On the other hand we have
\begin{equation*}
\begin{split}
&2\langle {\mathfrak J}_{e_1}, v_{\scriptscriptstyle{N,\lambda}}\rangle_{\scriptscriptstyle{N,T}} - \|v_{\scriptscriptstyle{N,\lambda}} \|^2_{1,\lambda} \\
&= 4a \langle {\mathfrak J}_{e_1}, {\mathfrak J}_{e_1}\rangle_{\scriptscriptstyle{N,T}} - 4a^{2} \langle {\mathfrak J}_{e_1}, (\lambda -\gamma {\mc S}_N){\mathfrak J}_{e_1} \rangle_{\scriptscriptstyle{N,T}} \\
&=4a \langle {\mathfrak J}_{e_1}, {\mathfrak J}_{e_1}\rangle_{\scriptscriptstyle{N,T}} - 4a^{2} (\lambda +\gamma) \langle {\mathfrak J}_{e_1},  {\mathfrak J}_{e_1} \rangle_{\scriptscriptstyle{N,T}} 
\end{split}
\end{equation*}
because ${\mc S}_N {\mathfrak J}_{e_1} = -{\mathfrak J}_{e_1}$. By a simple computation we have 
\begin{equation}
\label{eq:JJ}
 \langle {\mathfrak J}_{e_1}, {\mathfrak J}_{e_1} \rangle_{\scriptscriptstyle{N,T}}  = 4N T {\text{Var}}(V' (r_0))
\end{equation}
If $a$ is sufficiently small we get 
\begin{equation*}
 N^{-d} \langle {\mathfrak J}_{e_1}, (\lambda -{\mc L}_N)^{-1} {\mathfrak J}_{e_1} \rangle_{\scriptscriptstyle{N,T}}  \ge C^{-1}
\end{equation*}
with $C>0$ independent of $N$ and $\lambda$.
\end{proof}

\subsection{Disordered harmonic chain}
\label{subsec:DHC}

In this subsection we are interested in the effect of disorder on the thermal conductivity properties. The simplest way to introduce randomness is to assume that masses of atoms can vary from site to site according to a random sequence. We first review basic facts for deterministic chains with random masses. As it is well known, the presence of disorder generally induces localization of  the normal modes  and one can expect to have a perfect thermal insulators ($\kappa_N (T) \to 0$). The only analytically tractable model is the one dimensional disordered harmonic chain (DHC).  Surprisingly the behavior of the thermal conductivity depends on boundary conditions and on the properties of the thermostats (\cite{CL}, \cite{RG}). This curious phenomenon has been studied in \cite{D} (see also \cite{RD}) in a more general setting and it turns out that "the exponent [of $\kappa_N$] depends not only on the properties of the disordered chain itself, but also on the spectral properties of the heat baths. For special choices of baths one gets the "Fourier  behavior" ".  If we add a pinning potential in the DHC, $\kappa_N (T)$ becomes exponentially small in $N$.  

Recently, Dhar and Lebowitz (\cite{DL}) were interested in the effect of both disorder and anharmonicity. The conclusions of their numerical simulations are that the introduction of a small amount of phonon-phonon interactions in the DHC leads to a positive finite thermal conductivity.  

We consider now the $(\alpha,\nu)$-harmonic chain with random masses and energy conserving noise. The noise should simulate in some sense nonlinearity effects and  in view of the numerical simulations of \cite{DL} one would expect the model to become a normal conductor : $\kappa_N (T) \to \kappa (T)$ with $\kappa(T)$ finite and positive. We are not able to obtain interesting informations for $\kappa (T)$ but only for $\kappa^{GK} (T)$. Hence the behavior of the thermal conductivity is studied in the linear response theory framework by using the Green-Kubo formula. Behavior of the conductivity defined through Green-Kubo formula has not been studied for DHC. It would be interesting to know what is the order of divergence of the latter. For the perturbed DHC we obtain uniform finite positive lower and upper bounds for the $d$ dimensional finite volume Green-Kubo formula of the thermal conductivity with or without pinning (Theorem \ref{th:QGK}) so that the thermal conductivity is always finite and positive. In particular it shows the presence of the noise is sufficient to destroy localization of eigen-functions in pinned DHC. Linear response approach avoids the difficulty to deal with a nonequilibrium setting where effects of spectral properties of heat baths could add difficulties as in the case of purely DHC. In the nonequilibrium setting, we expect that since the Green-Kubo formula for the thermal conductivity of the perturbed DHC remains finite, it will not depend on the boundaries. 

\begin{theorem}{\cite{B2}}
\label{th:QGK}
There exists a positive constant $C>0$ independent of $\lambda$ and $N$ such that
\begin{equation*}
\begin{split}
C^{-1} \le \liminf_{\lambda \to 0} \liminf_{N \to \infty} \int_0^{\infty} e^{-\lambda t} N^{-d} {\mathbb E}_{\scriptscriptstyle{N,T}}{\left[{\frak J}_{e_1}(t),\, {\frak J}_{e_1} (0) \right]}dt \\
 \le \limsup_{\lambda \to 0} \limsup_{N\to \infty} \int_0^{\infty} e^{-\lambda t} N^{-d} {\mathbb E}_{\scriptscriptstyle{N,T}} {\left[{\frak J}_{e_1}(t),\, {\frak J}_{e_1} (0) \right]}dt \le C
\end{split}
\end{equation*}
\end{theorem}

\begin{proof}
The proof is similar to what is presented in subsection \ref{subsec:ac} for the anharmonic case. 
\end{proof}

A priori the Green-Kubo formula $\kappa^{GK}(T)$ depends on the particular realization of the masses $\{m_x\}$. But a formal argument (\cite{B2}) suggests that if masses are distributed according to some stationary ergodic probability measure ${\mathbb P}^*$ then $\kappa^{GK}(T)$ depends only on the statistics of the masses and not on the particular realization of the disorder. One can write formally an infinite Green-Kubo  homogenized formula $\kappa_{hom.}(T)$ obtained by averaging over the masses. In order to define $\kappa_{hom.} (T)$ one has to consider the infinite volume dynamics with generator ${\mc L}$ given by
\begin{equation*}
{\mc L} = \sum_{x \in \Z^d} \left\{ \partial_{p_x} \mathcal H \cdot \partial_{q_x} -
     \partial_{q_x}\mathcal H \cdot  \partial_{p_x} \right\} + \gamma \sum_{i,j,k=1}^d \sum_{x \in \Z^d} (X_{x,x+e_k}^{i,j})^2
\end{equation*}
and 
$$ {\mc H} = \sum_{x \in \Z^d} \cfrac{p_x^2}{m_x} + (\alpha \Delta -\nu) q_x \cdot q_x$$
Observe that the sums are taken over ${\Z}^d$. One can show that the dynamics with generator ${\mc L}$ is well defined. Then we have
\begin{equation*}
\kappa^{GK}_{hom.} (T)= \gamma/d +\lim_{\lambda \to 0} {\mathbb E}^* \left[  \int_0^\infty dt e^{-\lambda t} \sum_{z \in \mathbb Z^d} {\mathbb E}_{\scriptscriptstyle{T}} \left[ j^{a}_{0,e_1} (t) \tau_z j^{a}_{0,e_1}(0)\right] \right] 
\end{equation*}
where $\tau_z$ denotes the shift on the configuration space and ${\mathbb E}_{\scriptscriptstyle{T}}$ is the expectation corresponding to the infinite dynamics starting from the infinite volume Gibbs measure with temperature $T$.
 
\begin{theorem}{\cite{B2}}
\label{th:AGK}
Assume that $\{m_x\}$ is stationary under ${\mathbb P}^*$ and there are positive constants $\underline{m}$ and ${\overline m}$ such that
\begin{equation*}
{\mathbb P}^* ({\underline m} \leq m_x \leq {\overline m})=1
\end{equation*}
The Hamiltonian contribution to the homogenized Green-Kubo formula for the thermal conductivity $\kappa^{GK}_{hom.} (T)  -\gamma/d$
\begin{equation*}
\label{eq:GKinfty}
\kappa^{1,1}_{hom.}(T) - \gamma/d =\lim_{\lambda \to 0} {\mathbb E}^* \left[  \int_0^\infty dt e^{-\lambda t} \sum_{z \in \mathbb Z^d} {\mathbb E}_{\scriptscriptstyle{T}} \left[ j^{a}_{0,e_1} (t) \tau_z j^{a}_{0,e_1}(0)\right] \right] 
\end{equation*}
exists, is positive and finite.
 \end{theorem}

\begin{proof}
The proof is based on functional analysis arguments (\cite{Ben}, \cite{BLLO}, \cite{B2}).
\end{proof}

\section{Energy-momentum conserving model}
\label{sec:emc}
We investigate the same problems as in the previous section but for the chain perturbed by the energy-momentum conserving noise. Momentum is then conserved by the total dynamics if it is conserved by the Hamiltonian dynamics which is the case if and only if the system is unpinned. Dramatic consequences in low dimensional systems appear. In fact we prove that for unpinned systems with harmonic interactions, thermal conductivity is infinite in 1 and 2 dimensions, while is finite for $d\ge 3$ or for pinned systems. So thermal conductivity in our model behaves qualitatively like in a deterministic nonlinear system, i.e. these stochastic interactions reproduce some of the features of the nonlinear
deterministic hamiltonian interactions. 

For anharmonic systems, even with the stochastic noise, we are not able
to prove the existence of thermal conductivity (finite or
infinite). If the dimension $d$ is greater than $3$ and the system is
pinned, we get a uniform bound on the finite size system
conductivity. For low dimensional pinned systems ($d=1,2$), we can
show the conductivity is finite if the interaction potential is
quadratic and the pinning is generic. For the unpinned system we have
to assume that the interaction between nearest-neighbor particles is
strictly convex and quadratically bounded at infinity. This because we
need some informations on the spatial decay of correlations in the
stationary equilibrium measure, that decay slowly in unpinned system. In this case, we prove the conductivity is finite in dimension $d\geq 3$ and we obtain upper bounds in the size $N$ of the
system of the form $\sqrt{N}$ in $d=1$ and $(\log N)^2$ in $d=2$
(see Theorem \ref{th-anharm}).

\subsection{Homogenous harmonic chain}

Consider first the homogenous $(\alpha,\nu)$-harmonic case (\ref{eq:harm}) with all masses equal to $1$. In \cite{BBO1},\cite{BBO2}, we obtained the following theorems. 
\begin{theorem}{\cite{BBO2}}
\label{th-harm-cond}
In the $(\alpha,\nu)$-harmonic case (\ref{eq:harm}), the limit
defining $\kappa^{GK} (T)$ exists. It is finite if $d\geq 3$ or if the on-site harmonic potential is present
($\nu>0$), and is infinite in the other cases. 
When finite $\kappa^{GK} (T)$ is independent of $T$ and
the following formula holds  
\begin{equation*}\label{cond1}
{\kappa}= \cfrac{1}{8\pi^2 d \gamma}
\int_{[0,1]^d}\cfrac{(\partial_{{k}^1}\omega)^2(k)}{\psi
  (k)}d k + \cfrac{\gamma}{d}
\end{equation*}
where $\omega(k)$ is the dispertion relation
\begin{equation*}
\label{eq:dr}
\omega (k) = 
\left(\nu + 4\alpha\sum_{j=1}^{d} \sin^{2} (\pi k^j)\right)^{1/2}
\end{equation*}
and
\begin{equation*}
\label{eq:psi}
\psi (k) = 
\begin{cases}
8\sum_{j=1}^{d} \sin^{2} (\pi k^j), \quad \mbox{ if } \quad d\geq 2\\
4/3 \ \sin^2 (\pi k) (1+2\cos^2 (\pi k)), \quad \mbox{ if } \quad d=1
\end{cases}
\end{equation*}
\end{theorem}

Consequently in the unpinned harmonic cases in dimension $d=1$ and $2$, the conductivity of our model diverges. We have the following behavior for the truncated Green-Kubo formula.

\begin{theorem}{\cite{BBO2}}
  \label{th-ldharm}
In the harmonic case, if $W= 0$:
\begin{enumerate}
\item $\kappa^{GK}_N (T) \sim N^{1/2}$ if $d=1$,
\item $\kappa^{GK}_N (T) \sim \log N$ if $d=2$.
\end{enumerate}
In all other cases $\kappa^{GK}_N$ is bounded in $N$ and converges to
$\kappa^{GK} (T)$. \\
\end{theorem}

The interest of these theorems is that they show energy-momentum conserving model reproduces the expected behavior of purely Hamiltonian  nonlinear chains, meaning an infinite conductivity for low dimensional unpinned systems and a finite conductivity otherwise. Several microscopic stochastic models have been proposed in the past but none of them has this property with respect to momentum conservation and dimension.

The strategy of the proof is very similar to the proof given for the energy conserving model. Recall (\ref{eq:18}). Then the problem is reduced to solve the resolvent equation
\begin{equation*}
\lambda u_{\scriptscriptstyle{N,\lambda}} -{\mc L}_N u_{\scriptscriptstyle{N,\lambda}} ={\mathfrak J}_{e_1}
\end{equation*}
Solving explicitly such an equation is in general very difficult. The key property of ${\mc L}_N$ is that for quadratic potentials $V$ and $W$ it sends a polynomial function of $p$'s and $q$'s of degree $k$ to a polynomial function of the same degree. Since ${\mathfrak J}_{e_1}$ is of degree $2$, $u_{N,\lambda}$ has to be searched in the smaller space of polynomial functions of degree $2$. Moreover the translation invariance of ${\mathfrak J}_{e_1}$ implies that $u_{N,\lambda}$ is in the form
\begin{equation*}
u_{N,\lambda} =\sum_{x,y} f_{N,\lambda}(x-y)\, p_x \cdot p_y +\sum_{x,y} h_{N,\lambda} (x-y)\,  q_{x} \cdot q_{y} + \sum_{x,y} g_{N,\lambda}(x-y)\,  p_x \cdot q_y
\end{equation*}
where $f_{N,\lambda}, g_{\scriptscriptstyle{N,\lambda}}$ and $h_{\scriptscriptstyle{N,\lambda}}$ are functions from ${\To}_N^d$ to $\mathbb R$. Then we obtain linear equations for $f_{\scriptscriptstyle{N,\lambda}}, g_{\scriptscriptstyle{N,\lambda}}$ and $h_{\scriptscriptstyle{N,\lambda}}$. It turns out that $f_{N,\lambda}=h_{N,\lambda}=0$ and that $g_{N,\lambda}$ is simply related to the Green function of a symmetric simple random walk on $\To_N^d$. Then one computes easily 
\begin{equation*}
\langle u_{\scriptscriptstyle{N,\lambda}}, {\mathfrak J}_{e_1}\rangle_{\scriptscriptstyle{N,T}}
\end{equation*}
and we get the result.

\subsection{Homogenous anharmonic chain}

For the anharmonic chain we are not able to solve the resolvent equation
\begin{equation*}
\lambda  u_{\scriptscriptstyle{N,\lambda}}  -{\mc L}_N u_{\scriptscriptstyle{N,\lambda}} ={\mathfrak J}_{e_1}
\end{equation*}
Nevertheless we can use the same strategy as the one explained in subsection \ref{subsec:ac}. We introduce a variational formula to estimate by above
\begin{equation*}
\langle u_{\scriptscriptstyle{N,\lambda}}, {\mathfrak J}_{e_1}\rangle_{\scriptscriptstyle{N,T}}
\end{equation*}  

Extra assumptions on the potentials $V$ and $W$ assuring a uniform
control on the canonical static correlations have to be done. We have this control as soon as $V$ is strictly convex. In the pinned case $W>0$, this control is ``morally'' valid as soon as
the infinite volume Gibbs measure is unique.  Exact assumptions are given in \cite{BH}, theorem 3.1 and theorem 3.2. Hence "general anharmonic case" will refer to potentials $V$ and $W$ such
that this control is valid.

\begin{theorem}{\cite{BBO2}}
  \label{th-anharm}
Consider the general anharmonic case. There exists a constant $C$
(depending on the temperature $T$) such that 
 \begin{itemize}
\item For $d\ge 3$, 
  \begin{enumerate}\item 
    either $W_j>0$ is general
  \item or if $W_j=0$ and $0< c_-\le V_j''\le C_+< \infty$ for any
    $j$, 
  \end{enumerate}
 then 
$$
\kappa^{GK}_N (T) \le C.
$$

 \item For $d=2$, if $W_j=0$ and $0< c_-\le V_j''\le C_+< \infty$ for
   any $j$, then    
 $$
 \kappa^{GK}_N (T) \le C(\log N)^2.
 $$
 \item For $d=1$, if $W_j=0$ and $0< c_-\le V_j''\le C_+< \infty$, then 
 $$
 \kappa^{GK}_N (T) \le C {\sqrt N}.
 $$
\item
Moreover, in any dimension, if $V_j$ are quadratic and $W_j>0$ is
general then ${\kappa}^{GK}_N (T)\leq C$.  
\end{itemize}
\end{theorem}

\section{Open questions}
\label{sec:open}
In this section we mention several open questions. From the (probably) easier to the most difficult we have:
\begin{itemize}
\item Prove (\ref{eq:22}) in the $d$-dimensional linear pinned or unpinned case. It should be also valid in the nonlinear case.\\
\item  Consider the energy-momentum conserving model with linear interactions. We have studied the thermal conductivity in the Green-Kubo framework. Consider the same system but in contact with thermal baths at different temperatures. Can you say something about $\kappa_N (T)$ ? We expect it diverges like $\sqrt{N}$ (resp. $\log N$) for $d=1$ (resp. $d=2$) in the unpinned case and converges in the other cases to a finite positive constant. In fact a similar unpinned one-dimensional model has been considered in \cite{DLLP},\cite{LMP} and numerical and analytical (but not rigorous) results confirm this conjecture.  \\
\item Consider the $(\alpha,\nu)$-harmonic case with random masses and energy conserving noise. Can you say something about the temperature dependance of $\kappa_{hom.}^{GK}$ ? Can you prove the almost sure convergence (w.r.t. disorder) of the Green-Kubo formula to the homogenized Green-Kubo formula $\kappa_{hom.}^{GK}$ ?\\
\item Consider the $(\alpha,\nu)$-harmonic case with random masses and energy-momentum conserving noise. We have still upper bounds for the truncated Green-Kubo formula similar to the one obtained for the homogenous nonlinear chain in theorem \ref{th-anharm}.  Can you obtain lower bounds? In particular do you have a positive conductivity for pinned chains ?\\
\item Can you prove convergence of the Green-Kubo formula for nonlinear chains with energy conserving noise ?\\
\item Consider  an homogenous nonlinear chain with energy-momentum conserving noise. Can you obtain (non trivial) lower bounds for the truncated Green-Kubo formula ? Can you prove convergence of the Green-Kubo formula in dimension $d \ge 3$ or for pinned systems ? Interesting and surprising numerical simulations are provided in \cite{BDLLOP}.\\
\item Consider  an homogenous nonlinear chains with energy-momentum conserving noise in contact with two thermal baths at different temperatures. Can you say something about $\kappa_N (T)$ ?\\
\item Consider a nonlinear chain with energy-momentum conserving noise and random masses. Can you say anything about $\kappa (T)$ or $\kappa^{GK} (T)$ ? 
  
\end{itemize}

\begin{acknowledgement}
We acknowledge the support of the French Ministry of Education through the ANR BLAN07-2184264 grant.
\end{acknowledgement}

\input{referenc-bernardin}

\end{document}

%% file: referenc-bernardin.tex
%
%
%